\documentclass[11pt, a4paper, logo, copyright, nonumbering]{technical_report}
\usepackage[authoryear, sort&compress, round]{natbib}
\usepackage{dblfloatfix}
\usepackage{ulem}
\usepackage{caption}
\usepackage{dramatist}
\usepackage{xspace}
\usepackage{pifont} 
\usepackage{multirow}
\usepackage{tcolorbox}
\usepackage{xltabular}
\usepackage{longtable}
\usepackage{hyperref}
\usepackage{amsfonts}
\usepackage{amsmath}
\usepackage{amssymb}
\usepackage{lineno}
\usepackage{multirow}
\usepackage{adjustbox}

\usepackage[bottom]{footmisc}

\usepackage{CJKutf8}
\usepackage{subfigure}
\usepackage{setspace}

\usepackage{dsfont}
\usepackage{array} 
\usepackage{tabularx} 
\usepackage{subfigure} 
\usepackage{xcolor} 

\usepackage{lipsum}  
\usepackage{multicol} 

\makeatletter
\def\@BTrule[#1]{%
  \ifx\longtable\undefined
    \let\@BTswitch\@BTnormal
  \else\ifx\hline\LT@hline
    \nobreak
    \let\@BTswitch\@BLTrule
  \else
     \let\@BTswitch\@BTnormal
  \fi\fi
  \global\@thisrulewidth=#1\relax
  \ifnum\@thisruleclass=\tw@\vskip\@aboverulesep\else
  \ifnum\@lastruleclass=\z@\vskip\@aboverulesep\else
  \ifnum\@lastruleclass=\@ne\vskip\doublerulesep\fi\fi\fi
  \@BTswitch}
\makeatother

\addto\extrasenglish{
}

 {\begin{list}{}%
         {\setlength{\leftmargin}{#1}}%
         \item[]%
 }
 {\end{list}}
 
\reportnumber{001} 

\renewcommand{\today}{}

\title{\centering Technical Report: A Practical Guide to Kaldi ASR Optimization}

\author{
Mengze Hong \quad Di Jiang*
\\
\small
\texttt{mengze.hong@connect.polyu.hk \quad dijiang@webank.com}
}

\renewcommand{\phi}{\varphi}

\renewcommand{\epsilon}{\varepsilon}
\renewcommand{\imath}{\mathrm{i}}

\newlength{\restsubwidth}
\newlength{\restsubheight}
\newlength{\restsubmoreheight}
\newcommand{\rest}[2]{%
        \settowidth{\restsubwidth}{\ensuremath{#2}}
        \settoheight{\restsubheight}{\ensuremath{{}_{#2}}}
        \ensuremath{{#1\hskip 0.5pt}_{\vrule\kern2pt\parbox[b][%
        4pt][b]{\the\restsubwidth}{%
                        \ensuremath{{}_{#2}}}}}
        }

\begin{abstract}

This technical report introduces innovative optimizations for Kaldi-based Automatic Speech Recognition (ASR) systems, focusing on acoustic model enhancement, hyperparameter tuning, and language model efficiency. We developed a custom Conformer block integrated with a multistream TDNN-F structure, enabling superior feature extraction and temporal modeling. Our approach includes advanced data augmentation techniques and dynamic hyperparameter optimization to boost performance and reduce overfitting. Additionally, we propose robust strategies for language model management, employing Bayesian optimization and $n$-gram pruning to ensure relevance and computational efficiency. These systematic improvements significantly elevate ASR accuracy and robustness, outperforming existing methods and offering a scalable solution for diverse speech recognition scenarios. This report underscores the importance of strategic optimizations in maintaining Kaldi's adaptability and competitiveness in rapidly evolving technological landscapes.

\end{abstract}

\begin{document}
\begin{CJK*}{UTF8}{gbsn}

\maketitle




\begin{figure*}[h]
    \centering
    \includegraphics[width=\linewidth]{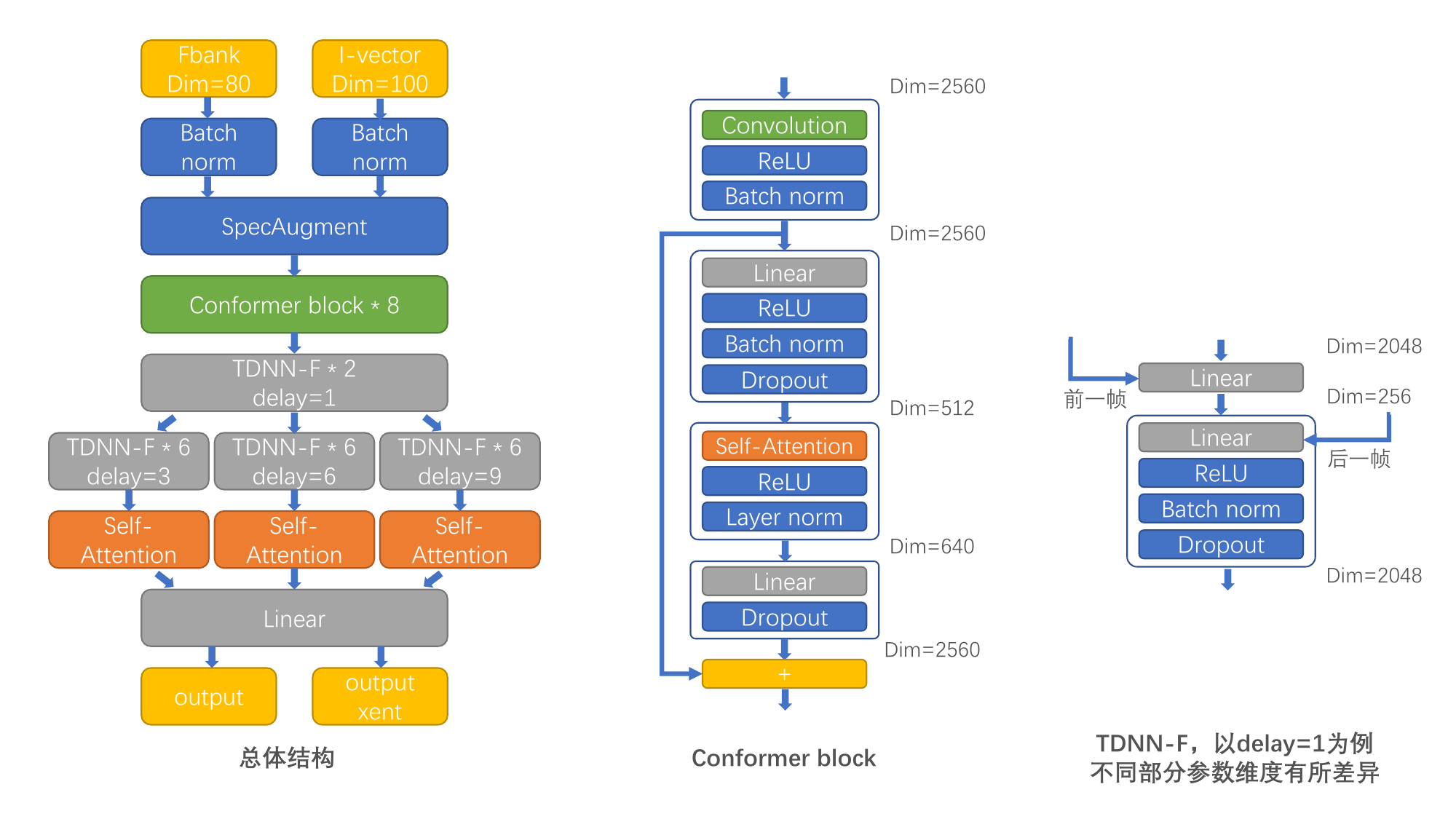}
    \caption{\centering Acoustic Model Architecture: structure overview (left), custom Conformer block (middle), and TDNN-F component (right)}
    \label{fig:am}
\end{figure*}

\newpage

\section{Introduction}

Automatic Speech Recognition (ASR), commonly known as speech-to-text technology, is designed to transcribe spoken language into written text by converting audio signals into sequences of words. Formally, let \(\mathbf{X} = \{x_1, x_2, \ldots, x_T\}\) represent the input sequence of acoustic features, where each \(x_t \in \mathbb{R}^d\) is a feature vector at time \(t\), and \(T\) denotes the length of the sequence. The objective of ASR is to determine the most probable sequence of words \(\mathbf{W} = \{w_1, w_2, \ldots, w_N\}\) that corresponds to the acoustic feature sequence \(\mathbf{X}\). This objective can be expressed as an optimization problem:
\[
\mathbf{W}^* = \arg\max_{\mathbf{W}} P(\mathbf{W} | \mathbf{X}),
\]
where \(P(\mathbf{W} | \mathbf{X})\) is the posterior probability of the word sequence given the acoustic input. Using Bayes' theorem, this probability is decomposed as:
\[
P(\mathbf{W} | \mathbf{X}) = \frac{P(\mathbf{X} | \mathbf{W}) P(\mathbf{W})}{P(\mathbf{X})},
\]
where \(P(\mathbf{X} | \mathbf{W})\) is the acoustic model probability, \(P(\mathbf{W})\) is the language model probability, and \(P(\mathbf{X})\) is the evidence, typically treated as a normalization constant during decoding. Modern ASR systems, particularly those leveraging deep learning, model \(P(\mathbf{X} | \mathbf{W})\) using neural networks such as Convolutional Neural Networks (CNNs), Recurrent Neural Networks (RNNs), or Transformers, and approximate \(P(\mathbf{W})\) using \(n\)-gram or neural language models.

The significance of ASR lies in its versatility across diverse applications, such as audio transcription for meetings and broadcasts \citep{adedeji2024sound, 10.1145/3474085.3478556, najafian2017automatic} and code-switching detection in bilingual speech \citep{diwan2021multilingual}. Additionally, ASR is essential for speaker diarization and verification \citep{mao2020speech, 9613759, Chen_Jiang_Peng_Lian_Zhang_Xu_Fan_Yang_2021}, keyword spotting \citep{michaely2017keyword}, and dialogue moderation \citep{10145599}, serving as critical components in enhancing safety and user experience for call centers. Among these, Kaldi \citep{povey2011kaldi} stands out as a pivotal open-source toolkit for ASR research and deployment, facilitating large-vocabulary continuous speech recognition \citep{povey2011kaldi}. Its modular architecture, support for diverse languages, and flexibility in handling complex speech tasks have made it a cornerstone in both academic and industrial settings.

The evolution of ASR, and Kaldi’s role within it, can be traced from early systems like Audrey \citep{furui1994overview}, through the era of hybrid Hidden Markov Model (HMM) and Gaussian Mixture Model (GMM) systems, and ultimately to the neural network era and beyond. Early ASR systems relied on HMM-GMM frameworks \citep{baker1975stochastic, rabiner2002tutorial}, which modeled the acoustic likelihood as:
\[
P(\mathbf{X} | \mathbf{W}) = \sum_{\mathbf{S}} \prod_{t=1}^T P(x_t | s_t) P(s_t | s_{t-1}),
\]
where \(\mathbf{S} = \{s_1, \ldots, s_T\}\) is a state sequence, \(P(x_t | s_t)\) is typically modeled by a GMM, and \(P(s_t | s_{t-1})\) represents state transition probabilities. These frameworks effectively captured acoustic and temporal patterns but struggled with complex speech variations due to their reliance on statistical modeling. The neural network era, beginning in the 2010s, marked a significant shift with the introduction of Deep Neural Networks (DNNs) \citep{hinton2012deep}. DNN-HMM hybrids \citep{dahl2012context} improved recognition accuracy by replacing GMMs with DNNs to estimate state posteriors, approximating the log-likelihood as:
\[
\log P(\mathbf{X} | \mathbf{W}) \approx \sum_{t=1}^T \log P(s_t | x_t; \theta_{\text{DNN}}) + \log P(s_t | s_{t-1}),
\]
where \(\theta_{\text{DNN}}\) denotes the parameters of the DNN. This approach significantly enhanced performance by leveraging the discriminative power of neural networks. \cite{seide2011conversational} scaled DNNs for conversational speech, further advancing performance. \cite{povey2018semi} introduced Time-Delay Neural Networks (TDNNs) for robust feature extraction. The integration of Connectionist Temporal Classification (CTC) \citep{graves2006ctc} and Recurrent Neural Network Transducers (RNN-T) \cite{graves2012sequence} further bridged hybrid and end-to-end approaches, with \cite{toshniwal2018endtoend} showcasing Kaldi’s adaptability to CTC-based systems.

As ASR progressed into the language model era, large-scale pre-trained models began to dominate. \cite{vaswani2017attention} introduced the Transformer architecture, revolutionizing sequence modeling and enabling end-to-end ASR systems like wav2vec 2.0 \citep{baevski2020wav2vec}. \cite{radford2023robust} further advanced this with Whisper, a multilingual end-to-end model that outperformed hybrid systems in low-resource settings. Concurrently, Kaldi-based research adapted to these trends. \cite{park2019specaugment} introduced SpecAugment, enhancing Kaldi’s robustness through data augmentation. \cite{zhu2020multilingual} explored cross-lingual transfer learning, improving Kaldi’s performance for under-resourced languages. Despite these advancements, optimizing Kaldi remains crucial. The computational demands of large model training necessitate efficient acceleration strategies. Addressing challenges in low-resource languages \citep{diwan2021lowresource} calls for innovative data augmentation and transfer learning approaches \citep{huang2021meta}. Furthermore, integrating Kaldi with contemporary architectures \citep{gulati2020conformer} is vital to stay competitive with end-to-end systems, necessitating comprehensive optimization in the model architecture and training process.

In this paper, we present a series of strategic advancements and optimizations tailored to enhance Kaldi ASR systems. Our main contributions are as follows:

\begin{enumerate}
    \item \textbf{Innovative Acoustic Model Components:} We designed a custom Conformer block that integrates convolution and self-attention, along with a multistream TDNN-F structure. This enhances feature extraction and temporal modeling within the Kaldi framework without significantly increasing computational requirements.
    
    \item \textbf{Acoustic Model Optimization:} We upgraded to 80-dimensional log Mel filterbank features and incorporated SpecAugment to improve performance and mitigate overfitting, especially in complex architectures.
    
    \item \textbf{Dynamic Hyperparameter Tuning:} We eliminated $\ell_2$ regularization and dynamically adjusted the cross-entropy weight during training to optimize performance and prevent model divergence.
    
    \item \textbf{Comprehensive Language Model Data Handling:} We established robust strategies for data selection, processing, and augmentation to ensure the relevance of training data and improve model robustness. This includes optimizing $n$-gram model thresholds and employing Bayesian optimization for model merging.
    
    \item \textbf{Efficient $n$-gram Model Training:} We optimized $n$-gram model training using KenLM for memory efficiency and applied SRILM pruning to reduce model size. We also leveraged perplexity and strategic keyword management to enhance evaluation efficiency and recall, minimizing loss in accuracy.
\end{enumerate}

\section{Acoustic Model Optimization}

We first introduce the basic architecture of an acoustic model built for Kaldi ASR, based on a typical CNN-TDNN-Attention architecture but optimized specifically for this task. Then, we present the hyperparameter optimization process with a hybrid loss.


\subsection{DNN Architecture}

The model architecture, depicted in Figure \ref{fig:am}, is a refined version of the CNN-TDNN-Attention structure \citep{miao2019new}. It incorporates 80-dimensional log Mel filterbank features and 100-dimensional i-vectors as input. Experimentation revealed that using 80-dimensional features yields better performance than the previously utilized 40-dimensional features. SpecAugment has been integrated into the model to effectively prevent overfitting, particularly when multistream structures are introduced, which tend to increase the risk of overfitting. Although the inclusion of SpecAugment slightly increases the training loss, it results in a lower test Character Error Rate (CER), achieving favorable outcomes even in the early training stages.

The Conformer block design draws inspiration from Google's Conformer architecture \citep{gulati2020conformer}. However, the original structure did not perform optimally within the Kaldi framework, possibly due to differences in the implementation of self-attention. Consequently, we adopted the concept of combining convolution and self-attention to design a \textbf{custom Conformer block}. In this structure, convolution is initially applied at the default dimension of 2560, followed by a linear layer that reduces the dimension to 512. This is then connected to multi-head self-attention, utilizing 8 heads, each with a value dimension of 80, resulting in an output dimension of 640. A final linear layer maps the dimension back to the original 2560, incorporating a skip-connection that adds the convolution output to the final result. The introduction of the initial linear layer serves to reduce dimensionality, lowering the computational demand of the self-attention component, while the latter linear layer ensures dimensional compatibility for skip-connection integration. By incorporating skip-connections, we address the challenge of training deep models, which significantly enhances performance. Unlike typical usage, this design does not bypass the convolution part for two reasons: firstly, convolution effectively extracts feature information \citep{pmlr-v260-hong25a}, which we aim to preserve; secondly, sub-sampling within some convolution operations may result in dimensional discrepancies between input and output.

The Factorized TDNN (TDNN-F) component retains a consistent design with the previous CNN-TDNN-Attention framework \citep{povey2018semi}, but introduces a multistream structure. This involves expanding the TDNN-F into three parallel streams, each with distinct delays, and concatenating their outputs. As the computational demand of TDNN-F is relatively low compared to the CNN component, adding two additional streams does not significantly increase overall computational load. Furthermore, skip-connections are implemented between different layers of TDNN-F.

\paragraph{Summary.} In the proposed model architecture, we introduced a custom Conformer block inspired by Google's architecture, combining convolution and self-attention for enhanced feature extraction, and expanded the TDNN-F component into three parallel streams with varying delays to improve temporal modeling. We upgraded input features from 40-dimensional to 80-dimensional log Mel filterbanks, which improved performance. SpecAugment was integrated to counteract overfitting, effectively reducing the test Character Error Rate (CER) in multistream structures despite an increase in training loss. We also implemented linear layers within the Conformer block to reduce dimensionality during self-attention operations, thus managing computational load, and incorporated skip-connections to enhance the training of deep models.

\subsection{Hyperparameter Optimization}

In the process of training, several hyperparameter optimizations were carried out to enhance the performance of the Kaldi ASR system:

The regularization coefficient (\texttt{chain.l2-regularize}) was set to zero. In the previous model configurations, $\ell_2$ regularization was applied to the parameters of the DNN. However, due to the strong regularization effect of SpecAugment, additional regularization did not yield better results. Instead, it led to model underfitting. Therefore, setting the regularization coefficient to zero proved to be more effective.

The weight of cross entropy (\texttt{chain.xent-regularize}) was dynamically adjusted during training. Kaldi's chain model employs a co-training strategy using both lattice-free maximal mutual information (LF-MMI) and cross-entropy (CE) losses. LF-MMI takes into account contextual sequence information and typically performs better than CE, although relying solely on LF-MMI can cause the model to exploit shortcuts and diverge from desired behaviors. Therefore, a combination of both losses is used, with the default CE weight set at 0.1. We observed that reducing the CE weight to 0.05 as training neared convergence further improved performance. However, lowering the CE weight too early in training could cause the model to diverge. For networks with Long Short-Term Memory (LSTM) structures, it is recommended to use a larger CE weight, such as 0.2, because LSTM networks optimize sequence-based losses like LF-MMI particularly quickly. This rapid optimization increases the risk of divergence, necessitating a stronger CE constraint to maintain stability.

In multilingual scenarios, such as those involving mixed Chinese and English speech, the number of context-dependent phone clusters was increased appropriately. The current training scripts limit the maximum number of CD phone clusters to 5000. In the case of purely Chinese phone clustering, over 4000 clusters were generated, indicating that the 5000 cluster limit is insufficient for mixed language scenarios. Increasing this limit prevents unrelated phones from being clustered together, thereby improving model performance in these contexts.

\paragraph{Summary.} In our hyperparameter optimization, we removed $\ell_2$ regularization from DNN parameters and instead leveraged SpecAugment's regularization to prevent underfitting. We dynamically adjusted the cross-entropy weight from 0.1 to 0.05 and recommended a higher weight (e.g., 0.2) for models with LSTM structures to offset rapid LF-MMI optimization. Additionally, we expanded the limit for context-dependent phone clusters beyond 5000 in multilingual scenarios to avoid clustering unrelated phones, thereby improving model accuracy in mixed-language contexts. These optimizations were critical in addressing specific challenges posed by the training data and the model architecture, improving the model accuracy and robustness.

\section{Language Model Optimization}

\subsection{Data Selection and Augmentation}

Strategic data selection is essential, focusing on relevance to the target scenario. For instance, when training models for live-stream ASR, it is important to limit the inclusion of telephone data to prevent it from dominating the training set. However, maintaining some scenario diversity is beneficial, as demonstrated by the observation that including a small amount of telephone data can enhance model performance. It is also advisable to exclude data that is excessively difficult to recognize, as it may cause alignment errors and lead to model deviation.

Data augmentation techniques, such as varying playback speed and random volume adjustments, are particularly effective in enhancing model training. While pitch shifting can also be beneficial, it does not yield additional improvements when paired with speed variations during acoustic model training, except for models sensitive to breathing sounds. Adding background noise is another powerful augmentation method, but it necessitates retraining i-vectors to achieve optimal results.

During training data processing, sentences containing rare characters or traditional Chinese characters not present in the Lexicon should be removed. For segmentation, forward maximum matching based on the current Lexicon is recommended \citep{zhao2018new, wong1996chinese}. Although general tokenization tools may offer more logical segmentation, their effectiveness diminishes if they are incompatible with the Lexicon. Efforts to develop an improved tokenizer based on the Lexicon have not shown significant improvements.

\paragraph{Summary.} Our approach focuses on strategic data selection and effective augmentation to improve model robustness and performance. We prioritize scenario-relevant data and employ techniques like playback speed and volume adjustments. Data processing involves filtering out incompatible characters and using forward maximum matching for segmentation to ensure Lexicon compatibility.

\subsection{$n$-gram Model Training}

An $n$-gram model estimates the probability of a word sequence $\mathbf{W} = \{w_1, w_2, \ldots, w_N\}$ by modeling the conditional probability of each word given its preceding $n-1$ words, playing a pivotal role in ASR by providing the language model probability $P(\mathbf{W})$ in the decoding process \citep{brown1992class, povey2011kaldi}. Formally, the probability of the sequence is approximated as:
\[
P(\mathbf{W}) = \prod_{i=1}^N P(w_i | w_{i-n+1}^{i-1}),
\]
where $w_{i-n+1}^{i-1} = \{w_{i-n+1}, \ldots, w_{i-1}\}$ represents the history of $n-1$ words, and the conditional probability is estimated from the corpus as:
\[
P(w_i | w_{i-n+1}^{i-1}) = \frac{C(w_{i-n+1}, \ldots, w_i)}{C(w_{i-n+1}, \ldots, w_{i-1})},
\]
where $C(\cdot)$ denotes the count of the specified $n$-gram or $(n-1)$-gram in the training corpus. In ASR, this probability model enhances the accuracy of transcribing acoustic features $\mathbf{X}$ into word sequences by weighting possible transcriptions during decoding, complementing the acoustic model $P(\mathbf{X} | \mathbf{W})$ to maximize the posterior probability $P(\mathbf{W} | \mathbf{X})$ in Kaldi-based systems.

The training of $n$-gram models requires carefully choosing threshold values, as experiments revealed that retaining all unigrams, along with bigrams and trigrams with a frequency greater than 3, yielded the best results on a 30GB corpus. In practice, it is advisable to set thresholds for bigrams and trigrams based on the corpus size, even for smaller datasets, to prevent the model from memorizing erroneous word pairings. An exception to this rule is with fixed expressions such as poetry, lyrics, or scripted dialogues. Since $n$-gram training is memory-intensive, especially with large corpora, KenLM is recommended for its ability to specify a maximum memory usage rate; if memory is insufficient, disk space can be used for temporary storage, albeit at a slower speed \cite{heafield-2011-kenlm}.

Given the large size of $n$-gram models, SRI Language Modeling (SRILM) can be used to prune them (command: \texttt{ngram -prune <threshold>}), with higher pruning thresholds resulting in smaller models \cite{stolcke2002srilm}. Experiments demonstrated that pruning notably reduces model size while minimally impacting CER. Training on large corpora and then pruning the model is more effective than training on smaller corpora, provided that the corpora are similar. Since pruning is relatively quick compared to training, it is advisable to initially train with less stringent thresholds to produce larger models, then prune using different thresholds to achieve a suitably sized model.

Regarding keyword management, adjusting their occurrence frequency can be done by duplicating or removing sentences containing keywords. However, caution is advised: excessive duplication can cause the $n$-gram to memorize fixed combinations, potentially triggering the next word incorrectly after preceding words. By incorporating an appropriate number of keywords into the training data, specifically one keyword per sentence, with duplicates based on word frequency, we can effectively improve keyword recall without negatively impacting accuracy. In practice, this method proved beneficial in the Audio Moderation System project from Tencent Tianyu, enhancing keyword recall without affecting accuracy, provided the number of keywords added is moderate.

Currently, experiments show that 4-gram models do not outperform 3-gram models. Due to their size, 5-gram models have only been tested on small-scale corpora, showing superior performance in some scenarios and inferior in others compared to 3-grams. Thus, 3-grams remain the preferred choice.

\paragraph{Summary.} In training $n$-gram models, threshold values for unigrams, bigrams, and trigrams are carefully chosen based on corpus size to prevent erroneous memorization, with KenLM recommended for efficient memory management. SRILM pruning significantly reduces model size while maintaining Character Error Rate (CER), allowing for the initial creation of larger models and subsequent threshold adjustments for optimal sizing. Keyword management is refined through strategic sentence duplication or removal, enhancing recall without diminishing accuracy. Experimentation shows 3-gram models as the preferred choice for their superior performance and manageable size compared to larger $n$-grams.

\subsection{Model Merging and Optimization}

Language model training should prioritize data relevance, especially for $n$-gram models, where the relevance of the training data significantly impacts performance \cite{wei2024acousticmodeloptimizationmultiple}. However, due to the limited availability of such data, it is often necessary to combine it with other datasets to ensure comprehensive coverage. Currently, the best practice is to train separate $n$-gram models on each relevant dataset and then merge them using a Bayesian optimization-based method. Let's denote the $n$-gram model trained on dataset $D_i$ as $M_i(n)$. If there are $k$ relevant datasets, we train $k$ separate $n$-gram models:
\[ M_1(n), M_2(n), \ldots, M_k(n). \]
Each model $M_i(n)$ is trained independently on its corresponding dataset $D_i$. The goal is to merge these models into a single model $M^*(n)$ that optimizes overall performance for the target scenario. This can be expressed as:
\[ M^*(n) = \arg\max_{M(n)} \mathcal{L}(M(n) \mid D_{\text{target}}), \]
where $\mathcal{L}(\cdot)$ represents the likelihood or performance metric on the target dataset $D_{\text{target}}$.

The merging process using a Bayesian optimization-based approach effectively transforms the task of constructing the optimal $n$-gram model into a model selection problem. The goal is to identify the optimal combination of model weights $\mathbf{w} = (w_1, w_2, \ldots, w_k)$, enabling the creation of a composite model that maximizes performance:
\[
M^*(n) = \sum_{i=1}^{k} w_i \cdot M_i(n), \quad \text{subject to} \quad \sum_{i=1}^{k} w_i = 1.
\]
Through Bayesian optimization, various weight configurations are evaluated iteratively, allowing for the exploration and exploitation of combinations that yield the highest performance on the target dataset. A text dataset from the target scenario, $D_{\text{target}}$, serves as a validation set during this merging process. Although a large dataset is not required, it should be representative enough to ensure that $M^*(n)$ generalizes well to the target scenario. This approach helps to create a more robust language model by leveraging the strengths of individual models trained on different datasets while optimizing for the specific needs of the target application.

The optimization target can also be enhanced to improve the training efficiency. Perplexity values on test sets are an effective metric for assessing language model performance. Our experiments indicate a positive correlation between transcript perplexity and final CER on the same test set, even though studies suggest perplexity and CER are not always correlated. As compiling the graph HCLG is time-consuming, optimizing the language model based on perplexity before compilation is advisable.

In contrast to the straightforward statistics of $n$-gram models, recurrent neural network language models (RNNLMs) offer feature abstraction capabilities \cite{tam2014asr, Lu2024}, making them more versatile and less demanding in terms of training data size. The most effective RNNLM is trained on a combination of live-stream transcripts and a 2GB sample of other corpora, which balances colloquial specificity with general applicability. This model excels across various scenarios, typically eliminating the need for retraining unless there is a Lexicon mismatch.

\paragraph{Summary.} Prioritizing data relevance and combining datasets using Bayesian optimization ensures comprehensive coverage and enhanced $n$-gram model performance. Perplexity values are utilized as an optimization metric before the costly HCLG compilation, leveraging their correlation with CER to improve training efficiency. Recurrent neural network language models (RNNLMs) provide versatility through feature abstraction, trained on a mix of live-stream transcripts and diverse corpora to balance specificity with general applicability, excelling across various scenarios with minimal retraining needs.



\section{Conclusion}

In this technical report, we have detailed a comprehensive approach to optimizing Kaldi-based Automatic Speech Recognition systems through innovations in acoustic model design, precise hyperparameter tuning, and strategic language model management. By integrating a custom Conformer block and multistream TDNN-F structure, our model achieves enhanced feature extraction and temporal modeling. We deployed advanced data augmentation techniques and dynamic hyperparameter adjustments to improve performance and mitigate overfitting. Our language model strategies, which include Bayesian optimization for merging $n$-gram models and the use of KenLM and SRILM for efficient training and pruning, ensure relevance and computational efficiency. These contributions collectively demonstrate significant enhancements in ASR accuracy and robustness, paving the way for future innovations in speech recognition technology. Our work underscores the importance of systematic optimizations in bolstering Kaldi's adaptability and competitiveness in diverse and evolving applications.

Future research and development should prioritize addressing data privacy constraints, which are increasingly important in today's technological landscape \citep{10.1145/3503161.3547731}. This will necessitate the adoption of advanced approaches such as transfer learning and federated learning to the conventioanl centralized ASR training \citep{10.1145/3447687}. Moreover, the integration of Large Language Models (LLMs) has shown substantial promise in enhancing traditional ASR system. Notably, LLMs can be employed for post-correction, refining transcription accuracy by dynamically adjusting outputs based on contextual understanding \citep{chen2023hyporadise}. Researchers have identified and categorized various methods within the LLM-in-the-loop framework \citep{Hong_2025}, which not only offer significant improvements in model performance but also present exciting new research opportunities in the rapidly evolving landscape of LLM applications, paving the way for future advancements in ASR technology.
\bibliography{main}

\begin{thebibliography}{42}
\providecommand{\natexlab}[1]{#1}
\providecommand{\url}[1]{\texttt{#1}}
\expandafter\ifx\csname urlstyle\endcsname\relax
  \providecommand{\doi}[1]{doi: #1}\else
  \providecommand{\doi}{doi: \begingroup \urlstyle{rm}\Url}\fi

\bibitem[Adedeji et~al.(2024)Adedeji, Joshi, and Doohan]{adedeji2024sound}
A.~Adedeji, S.~Joshi, and B.~Doohan.
\newblock The sound of healthcare: Improving medical transcription asr accuracy with large language models.
\newblock \emph{arXiv preprint arXiv:2402.07658}, 2024.

\bibitem[Baevski et~al.(2020)Baevski, Zhou, Mohamed, and Auli]{baevski2020wav2vec}
A.~Baevski, Y.~Zhou, A.~Mohamed, and M.~Auli.
\newblock wav2vec 2.0: A framework for self-supervised learning of speech representations.
\newblock \emph{Advances in neural information processing systems}, 33:\penalty0 12449--12460, 2020.

\bibitem[Baker(1975)]{baker1975stochastic}
J.~K. Baker.
\newblock The dragon system - an overview.
\newblock \emph{IEEE Transactions on Acoustics, Speech, and Signal Processing}, 23\penalty0 (1):\penalty0 24--29, 1975.

\bibitem[Brown et~al.(1992)Brown, Della~Pietra, Desouza, Lai, and Mercer]{brown1992class}
P.~F. Brown, V.~J. Della~Pietra, P.~V. Desouza, J.~C. Lai, and R.~L. Mercer.
\newblock Class-based n-gram models of natural language.
\newblock \emph{Computational linguistics}, 18\penalty0 (4):\penalty0 467--480, 1992.

\bibitem[Chen et~al.(2021)Chen, Jiang, Peng, Lian, Zhang, Xu, Fan, and Yang]{Chen_Jiang_Peng_Lian_Zhang_Xu_Fan_Yang_2021}
C.~Chen, D.~Jiang, J.~Peng, R.~Lian, C.~J. Zhang, Q.~Xu, L.~Fan, and Q.~Yang.
\newblock A health-friendly speaker verification system supporting mask wearing.
\newblock \emph{Proceedings of the AAAI Conference on Artificial Intelligence}, 35\penalty0 (18):\penalty0 16004--16006, May 2021.
\newblock \doi{10.1609/aaai.v35i18.17994}.
\newblock URL \url{https://ojs.aaai.org/index.php/AAAI/article/view/17994}.

\bibitem[Chen et~al.(2023{\natexlab{a}})Chen, Hu, Yang, Siniscalchi, Chen, and Chng]{chen2023hyporadise}
C.~Chen, Y.~Hu, C.-H.~H. Yang, S.~M. Siniscalchi, P.-Y. Chen, and E.-S. Chng.
\newblock Hyporadise: An open baseline for generative speech recognition with large language models.
\newblock \emph{Advances in Neural Information Processing Systems}, 36:\penalty0 31665--31688, 2023{\natexlab{a}}.

\bibitem[Chen et~al.(2023{\natexlab{b}})Chen, Jiang, Peng, Lian, Li, Zhang, Chen, and Fan]{9613759}
C.~Chen, D.~Jiang, J.~Peng, R.~Lian, Y.~Li, C.~Zhang, L.~Chen, and L.~Fan.
\newblock Scalable identity-oriented speech retrieval.
\newblock \emph{IEEE Transactions on Knowledge and Data Engineering}, 35\penalty0 (3):\penalty0 3261--3265, 2023{\natexlab{b}}.
\newblock \doi{10.1109/TKDE.2021.3127520}.

\bibitem[Chen et~al.(2024)Chen, Jiang, Tan, Song, Zhang, and Chen]{10145599}
Y.~Chen, D.~Jiang, C.~Tan, Y.~Song, C.~Zhang, and L.~Chen.
\newblock Neural moderation of asmr erotica content in social networks.
\newblock \emph{IEEE Transactions on Knowledge and Data Engineering}, 36\penalty0 (1):\penalty0 275--280, 2024.
\newblock \doi{10.1109/TKDE.2023.3283501}.

\bibitem[Dahl et~al.(2012)Dahl, Yu, Deng, and Acero]{dahl2012context}
G.~E. Dahl, D.~Yu, L.~Deng, and A.~Acero.
\newblock Context-dependent pre-trained deep neural networks for large-vocabulary speech recognition.
\newblock In \emph{IEEE International Conference on Acoustics, Speech and Signal Processing (ICASSP)}, pages 4773--4776, 2012.

\bibitem[Diwan et~al.(2021{\natexlab{a}})Diwan, Sitaram, and Choudhury]{diwan2021lowresource}
A.~Diwan, S.~Sitaram, and M.~Choudhury.
\newblock Asr for low-resource indian languages: A multilingual perspective.
\newblock \emph{ACM Transactions on Asian and Low-Resource Language Information Processing}, 20\penalty0 (6):\penalty0 1--24, 2021{\natexlab{a}}.

\bibitem[Diwan et~al.(2021{\natexlab{b}})Diwan, Vaideeswaran, Shah, Singh, Raghavan, Khare, Unni, Vyas, Rajpuria, Yarra, et~al.]{diwan2021multilingual}
A.~Diwan, R.~Vaideeswaran, S.~Shah, A.~Singh, S.~Raghavan, S.~Khare, V.~Unni, S.~Vyas, A.~Rajpuria, C.~Yarra, et~al.
\newblock Multilingual and code-switching asr challenges for low resource indian languages.
\newblock \emph{arXiv preprint arXiv:2104.00235}, 2021{\natexlab{b}}.

\bibitem[Furui(1996)]{furui1994overview}
S.~Furui.
\newblock An overview of speaker recognition technology.
\newblock \emph{Automatic Speech and Speaker Recognition: Advanced Topics}, pages 31--56, 1996.

\bibitem[Graves(2012)]{graves2012sequence}
A.~Graves.
\newblock Sequence transduction with recurrent neural networks.
\newblock In \emph{Neural Information Processing Systems (NIPS)}, 2012.

\bibitem[Graves et~al.(2006)Graves, Fernández, Gomez, and Schmidhuber]{graves2006ctc}
A.~Graves, S.~Fernández, F.~Gomez, and J.~Schmidhuber.
\newblock Connectionist temporal classification: Labelling unsegmented sequence data with recurrent neural networks.
\newblock In \emph{International Conference on Machine Learning (ICML)}, pages 369--376, 2006.

\bibitem[Gulati et~al.(2020)Gulati, Qin, Chiu, Parmar, Zhang, Yu, Han, Wang, Zhang, Wu, and Pang]{gulati2020conformer}
A.~Gulati, J.~Qin, C.-C. Chiu, N.~Parmar, Y.~Zhang, J.~Yu, W.~Han, S.~Wang, Z.~Zhang, Y.~Wu, and R.~Pang.
\newblock Conformer: Convolution-augmented transformer for speech recognition.
\newblock In \emph{INTERSPEECH}, pages 5036--5040, 2020.

\bibitem[Heafield(2011)]{heafield-2011-kenlm}
K.~Heafield.
\newblock {K}en{LM}: Faster and smaller language model queries.
\newblock In C.~Callison-Burch, P.~Koehn, C.~Monz, and O.~F. Zaidan, editors, \emph{Proceedings of the Sixth Workshop on Statistical Machine Translation}, pages 187--197, Edinburgh, Scotland, July 2011. Association for Computational Linguistics.
\newblock URL \url{https://aclanthology.org/W11-2123/}.

\bibitem[Hinton et~al.(2012)Hinton, Deng, Yu, Dahl, Mohamed, Jaitly, Senior, Vanhoucke, Nguyen, Sainath, and Kingsbury]{hinton2012deep}
G.~Hinton, L.~Deng, D.~Yu, G.~E. Dahl, A.-r. Mohamed, N.~Jaitly, A.~Senior, V.~Vanhoucke, P.~Nguyen, T.~N. Sainath, and B.~Kingsbury.
\newblock Deep neural networks for acoustic modeling in speech recognition: The shared views of four research groups.
\newblock \emph{IEEE Signal Processing Magazine}, 29\penalty0 (6):\penalty0 82--97, 2012.

\bibitem[Hong et~al.(2025{\natexlab{a}})Hong, Ng, Zhang, Wang, Song, and Jiang]{Hong_2025}
M.~Hong, W.~Ng, C.~J. Zhang, Y.~Wang, Y.~Song, and D.~Jiang.
\newblock Llm-in-the-loop: Replicating human insight with llms for better machine learning applications, May 2025{\natexlab{a}}.
\newblock URL \url{http://dx.doi.org/10.36227/techrxiv.174495034.42657551/v2}.

\bibitem[Hong et~al.(2025{\natexlab{b}})Hong, Zhang, Yang, SONG, and Jiang]{pmlr-v260-hong25a}
M.~Hong, C.~J. Zhang, L.~Yang, Y.~SONG, and D.~Jiang.
\newblock {InfantCryNet}: {A} data-driven framework for intelligent analysis of infant cries.
\newblock In V.~Nguyen and H.-T. Lin, editors, \emph{Proceedings of the 16th Asian Conference on Machine Learning}, volume 260 of \emph{Proceedings of Machine Learning Research}, pages 845--857. PMLR, 05--08 Dec 2025{\natexlab{b}}.
\newblock URL \url{https://proceedings.mlr.press/v260/hong25a.html}.

\bibitem[Huang et~al.(2021)Huang, Li, and Gong]{huang2021meta}
J.~Huang, J.~Li, and Y.~Gong.
\newblock Meta-learning for low-resource speech recognition.
\newblock In \emph{International Conference on Machine Learning (ICML)}, pages 4295--4304, 2021.

\bibitem[Jiang et~al.(2021)Jiang, Tan, Peng, Chen, Wu, Zhao, Song, Tong, Liu, Xu, Yang, and Deng]{10.1145/3447687}
D.~Jiang, C.~Tan, J.~Peng, C.~Chen, X.~Wu, W.~Zhao, Y.~Song, Y.~Tong, C.~Liu, Q.~Xu, Q.~Yang, and L.~Deng.
\newblock A gdpr-compliant ecosystem for speech recognition with transfer, federated, and evolutionary learning.
\newblock \emph{ACM Trans. Intell. Syst. Technol.}, 12\penalty0 (3), May 2021.
\newblock ISSN 2157-6904.
\newblock \doi{10.1145/3447687}.
\newblock URL \url{https://doi.org/10.1145/3447687}.

\bibitem[Lu et~al.(2024)Lu, Lian, Jiang, Song, Su, Wei, and Yang]{Lu2024}
J.~Lu, R.~Lian, D.~Jiang, Y.~Song, Z.~Su, V.~J. Wei, and L.~Yang.
\newblock Pretraining enhanced rnn transducer.
\newblock \emph{CAAI Artificial Intelligence Research}, 3:\penalty0 9150039, 2024.
\newblock \doi{10.26599/AIR.2024.9150039}.
\newblock URL \url{https://www.sciopen.com/article/10.26599/AIR.2024.9150039}.

\bibitem[Mao et~al.(2020)Mao, Li, McAuley, and Cottrell]{mao2020speech}
H.~H. Mao, S.~Li, J.~McAuley, and G.~Cottrell.
\newblock Speech recognition and multi-speaker diarization of long conversations.
\newblock \emph{arXiv preprint arXiv:2005.08072}, 2020.

\bibitem[Miao et~al.(2019)Miao, McLoughlin, and Yan]{miao2019new}
X.~Miao, I.~McLoughlin, and Y.~Yan.
\newblock A new time-frequency attention mechanism for tdnn and cnn-lstm-tdnn, with application to language identification.
\newblock In \emph{Interspeech}, pages 4080--4084, 2019.

\bibitem[Michaely et~al.(2017)Michaely, Zhang, Simko, Parada, and Aleksic]{michaely2017keyword}
A.~H. Michaely, X.~Zhang, G.~Simko, C.~Parada, and P.~Aleksic.
\newblock Keyword spotting for google assistant using contextual speech recognition.
\newblock In \emph{2017 IEEE Automatic Speech Recognition and Understanding Workshop (ASRU)}, pages 272--278. IEEE, 2017.

\bibitem[Najafian et~al.(2017)Najafian, Hsu, Ali, and Glass]{najafian2017automatic}
M.~Najafian, W.-N. Hsu, A.~Ali, and J.~Glass.
\newblock Automatic speech recognition of arabic multi-genre broadcast media.
\newblock In \emph{2017 IEEE Automatic Speech Recognition and Understanding Workshop (ASRU)}, pages 353--359. IEEE, 2017.

\bibitem[Park et~al.(2019)Park, Chan, Zhang, Chiu, Zoph, Cubuk, and Le]{park2019specaugment}
D.~S. Park, W.~Chan, Y.~Zhang, C.-C. Chiu, B.~Zoph, E.~D. Cubuk, and Q.~V. Le.
\newblock Specaugment: A simple data augmentation method for automatic speech recognition.
\newblock \emph{arXiv preprint arXiv:1904.08779}, 2019.

\bibitem[Povey et~al.(2011)Povey, Ghoshal, Boulianne, Burget, Glembek, Goel, Hannemann, Motlicek, Qian, Schwarz, Silovsky, Stemmer, and Vesely]{povey2011kaldi}
D.~Povey, A.~Ghoshal, G.~Boulianne, L.~Burget, O.~Glembek, N.~Goel, M.~Hannemann, P.~Motlicek, Y.~Qian, P.~Schwarz, J.~Silovsky, G.~Stemmer, and K.~Vesely.
\newblock The kaldi speech recognition toolkit.
\newblock In \emph{IEEE Automatic Speech Recognition and Understanding Workshop (ASRU)}, 2011.

\bibitem[Povey et~al.(2018)Povey, Cheng, Wang, Li, Xu, Yarmohammadi, and Khudanpur]{povey2018semi}
D.~Povey, G.~Cheng, Y.~Wang, K.~Li, H.~Xu, M.~Yarmohammadi, and S.~Khudanpur.
\newblock Semi-orthogonal low-rank matrix factorization for deep neural networks.
\newblock In \emph{Interspeech}, pages 3743--3747, 2018.

\bibitem[Rabiner(2002)]{rabiner2002tutorial}
L.~R. Rabiner.
\newblock A tutorial on hidden markov models and selected applications in speech recognition.
\newblock \emph{Proceedings of the IEEE}, 77\penalty0 (2):\penalty0 257--286, 2002.

\bibitem[Radford et~al.(2023)Radford, Kim, Xu, Brockman, McLeavey, and Sutskever]{radford2023robust}
A.~Radford, J.~W. Kim, T.~Xu, G.~Brockman, C.~McLeavey, and I.~Sutskever.
\newblock Robust speech recognition via large-scale weak supervision.
\newblock In \emph{International conference on machine learning}, pages 28492--28518. PMLR, 2023.

\bibitem[Seide et~al.(2011)Seide, Li, and Yu]{seide2011conversational}
F.~Seide, G.~Li, and D.~Yu.
\newblock Conversational speech transcription using context-dependent deep neural networks.
\newblock In \emph{INTERSPEECH}, pages 437--440, 2011.

\bibitem[Song et~al.(2021)Song, Jiang, Zhao, Huang, Xu, Wong, and Yang]{10.1145/3474085.3478556}
Y.~Song, D.~Jiang, X.~Zhao, X.~Huang, Q.~Xu, R.~C.-W. Wong, and Q.~Yang.
\newblock Smartmeeting: Automatic meeting transcription and summarization for in-person conversations.
\newblock In \emph{Proceedings of the 29th ACM International Conference on Multimedia}, MM '21, page 2777–2779, New York, NY, USA, 2021. Association for Computing Machinery.
\newblock ISBN 9781450386517.
\newblock \doi{10.1145/3474085.3478556}.
\newblock URL \url{https://doi.org/10.1145/3474085.3478556}.

\bibitem[Song et~al.(2022)Song, Lian, Chen, Jiang, Zhao, Tan, Xu, and Wong]{10.1145/3503161.3547731}
Y.~Song, R.~Lian, Y.~Chen, D.~Jiang, X.~Zhao, C.~Tan, Q.~Xu, and R.~C.-W. Wong.
\newblock A platform for deploying the tfe ecosystem of automatic speech recognition.
\newblock In \emph{Proceedings of the 30th ACM International Conference on Multimedia}, MM '22, page 6952–6954, New York, NY, USA, 2022. Association for Computing Machinery.
\newblock ISBN 9781450392037.
\newblock \doi{10.1145/3503161.3547731}.
\newblock URL \url{https://doi.org/10.1145/3503161.3547731}.

\bibitem[Stolcke et~al.(2002)]{stolcke2002srilm}
A.~Stolcke et~al.
\newblock Srilm-an extensible language modeling toolkit.
\newblock In \emph{Interspeech}, volume 2002, page 2002, 2002.

\bibitem[Tam et~al.(2014)Tam, Lei, Zheng, and Wang]{tam2014asr}
Y.-C. Tam, Y.~Lei, J.~Zheng, and W.~Wang.
\newblock Asr error detection using recurrent neural network language model and complementary asr.
\newblock In \emph{2014 IEEE International Conference on Acoustics, Speech and Signal Processing (ICASSP)}, pages 2312--2316. IEEE, 2014.

\bibitem[Toshniwal et~al.(2018)Toshniwal, Liao, Sak, and Livescu]{toshniwal2018endtoend}
S.~Toshniwal, H.~Liao, H.~Sak, and K.~Livescu.
\newblock End-to-end multi-talker speech recognition.
\newblock In \emph{INTERSPEECH}, pages 2222--2226, 2018.

\bibitem[Vaswani et~al.(2017)Vaswani, Shazeer, Parmar, Uszoreit, Jones, Gomez, Kaiser, and Polosukhin]{vaswani2017attention}
A.~Vaswani, N.~Shazeer, N.~Parmar, J.~Uszoreit, L.~Jones, A.~N. Gomez, L.~Kaiser, and I.~Polosukhin.
\newblock Attention is all you need.
\newblock In \emph{Neural Information Processing Systems (NeurIPS)}, pages 5998--6008, 2017.

\bibitem[Wei et~al.(2024)Wei, Wang, Jiang, Tan, and Lian]{wei2024acousticmodeloptimizationmultiple}
V.~J. Wei, W.~Wang, D.~Jiang, C.~Tan, and R.~Lian.
\newblock Acoustic model optimization over multiple data sources: Merging and valuation, 2024.
\newblock URL \url{https://arxiv.org/abs/2410.15620}.

\bibitem[Wong and Chan(1996)]{wong1996chinese}
P.-k. Wong and C.~Chan.
\newblock Chinese word segmentation based on maximum matching and word binding force.
\newblock In \emph{COLING 1996 Volume 1: The 16th International Conference on Computational Linguistics}, 1996.

\bibitem[Zhao et~al.(2018)Zhao, Li, Yin, and Sun]{zhao2018new}
Y.~Zhao, H.~Li, S.~Yin, and Y.~Sun.
\newblock A new chinese word segmentation method based on maximum matching.
\newblock \emph{J. Inf. Hiding Multim. Signal Process.}, 9\penalty0 (6):\penalty0 1528--1535, 2018.

\bibitem[Zhu et~al.(2020)Zhu, Haghani, Tripathi, Ramabhadran, Farris, Xu, Lu, Sak, Leal, Gaur, et~al.]{zhu2020multilingual}
Y.~Zhu, P.~Haghani, A.~Tripathi, B.~Ramabhadran, B.~Farris, H.~Xu, H.~Lu, H.~Sak, I.~Leal, N.~Gaur, et~al.
\newblock Multilingual speech recognition with self-attention structured parameterization.
\newblock In \emph{INTERSPEECH}, pages 4741--4745, 2020.

\end{thebibliography}

\end{CJK*}
\end{document}